\documentstyle[bezier]{article}
\begin{document}
\begin{flushright} UMD PP 95-88 \\ gr-qc/9501023
\end{flushright}
\centerline{\Large\bf Black Hole Collisions, Analytic Continuation,}
\vspace{2truemm}
\centerline{\Large\bf and Cosmic Censorship$^*$}
\bigskip \medskip
\centerline{\large Dieter R. Brill$^{\footnotesize 1}$}
\bigskip
\centerline{$^{\footnotesize 1}$
\small Department of Physics, University of Maryland,
College Park, MD 20742, USA}
\bigskip

{\small $^*$Lectures at First Samos Meeting on Cosmology Geometry and
Relativity.

\ To appear in Springer Lecture Notes in Physics.}

\vspace{1truecm} \noindent
{\bf Abstract:} Exact solutions of the Einstein-Maxwell equations that
describe moving black holes in a cosmological setting are discussed with
the aim of discovering the global structure and testing cosmic censorship.
Continuation beyond the horizons present in these solutions is necessary
in order to identify the global structure. Therefore the possibilities
and methods of analytic extension of geometries are briefly reviewed.
The global structure of the Reissner-Nordstr\"om-de Sitter geometry is
found by these methods. When several black holes are present, the exact
solution is no longer everywhere analytic, but less smooth extensions
satisfying the Einstein equations everywhere are possible. Some of these
provide counterexamples to cosmic censorship.

\vspace {.5truecm} \noindent
{\large\bf 1 Introduction}

\bigskip \noindent
The recently
discovered \cite{KT}
exact {\it dynamical} solution of Einstein's equations provides
a common thread that ties together three items of the title. This solution
describes a cosmology with several charged black holes in motion and
capable of collision.
To discover the details of the collision one needs to continue the
solution beyond the region in which it was originally defined. The
spacetime so continued can then contain a naked singularity and provide a
counterexample to some versions of the cosmic censorship hypothesis.

Because interesting applications of Einstein's equations frequently occur
in manifolds of complicated topology, which cannot be covered by a
single coordinate patch, the region in which a typical exact solution
is first known is often incomplete. The simplest way to complete it,
if possible, is by analytic extension. It is remarkable how little is
known in a systematic way about this frequently encountered problem.
Here we will not materially improve on this situation, but merely recall
some of what is known about analytic continuation, and discuss the most
common class of geometries for which a method exists.

Although it is tempting to expect that such a method can also analytically
continue the colliding
black holes of interest here, we will find that, surprisingly, these holes
are in general not everywhere analytic. There is of course nothing
unphysical about such behavior; it is what one expects in the presence of a
gravitational waves pulse. We exhibit continuations that are as smooth as
possible across the associated Cauchy horizon.  The naked singularities
of interest for cosmic censorship are then found on the other side
of such horizons. (Other horizons are cosmological and do not hide
singularities, but an ``antipodal'' part of the universe.)

\bigskip \bigskip \noindent
{\large\bf 2 Analytic Continuation of Spacetimes}

\bigskip \noindent
The purpose of spacetime, as originally conceived, is to describe the history
of all inertial observers. A (timelike) geodesically incomplete spacetime fails
to do this, so it behooves us to extend it as far as possible. (If even the
maximal extension is incomplete we can begin to ask questions about cosmic
censorship.) An analytic extension, if possible, is
preferred because of its uniqueness and ``permanence" (i.e., the continuation
of Einstein's equations is automatic).

Actually, if not only the metric but also the manifold needs to be continued,
then the continuation is not necessarily unique. A simple example is a finite
part of (flat) Minkowski space, which can be continued either to the complete
Minkowski space, or to one of the several locally Minkowskian spaces, such
as the torus. More generally, any simply connected part of spacetime that
can be continued to a multiply connected one can also be continued to a
covering of the latter. We will encounter such ambiguities in the
cosmological black hole case below. A somewhat more subtle example is the
Taub-Nut space, which has two distinct and inequivalent analytic extensions
\cite{TN}. In these ambiguous cases one needs to decide on such properties
as the topology of the extended manifold along with the extension of the
metric. Once the whole (smooth) manifold is known, its analytic structure is
essentially unique. Because our manifolds are real, the relevant notion is
real analyticity --- existence of local power series expansions.
Analyticity of the manifold means that the coordinate transformations
between neighborhoods are real analytic functions.

In this context the analytic continuation of functions, such as the metric
coefficients, is then unique, and the continuation satisfies the continued
differential equation if the latter is itself analytic,
as in the case of Einstein's equations.\footnote{\small Analyticity
is not assured, nor is it necessarily
to be expected on physical grounds, if there are source terms present.
Well-known examples are stellar models that are non-analytic on the
stellar surface.} To know this is, however, of little help in practical
problems where the metric is given in some coordinates, and we desire to
extend across a boundary where both the metric coefficients and the coordinates
are non-analytic. There appears to be no systematic criterion for deciding
whether analytic continuation is possible, and one usually has to rely on
ingenuity to find suitable new coordinates in which the metric is
analytic in the relevant region.

It can happen that neither the metric nor the coordinates are analytic
functions on the manifold, but the metric coefficients are analytic
functions of the coordinates; or they many be extendable to another (real)
range of the coordinates by an excursion in the complex plane. Examples
of this are found in the Schwarzschild metric at $r = 2m\/$ and $r = 0$
respectively. In either case we obtain solutions of the Einstein equations
in the new coordinate range, but it is a separate question whether and
how the geometry so described fits together with the original geometry,
and if so, whether the fit is analytic. Finding a proper overlap seems
to be the only way to assure the latter.\footnote{\small It is remarkable that
some solutions known in closed form (which is sometimes --- loosely --- called
``analytic") can be extended with a high degree of differentiability
(e.g. $C^{122}$ as in \cite{Cru}) but not analytically.} For a class
of metrics to be discussed below, of which the Schwarzschild metric
is a member, one knows how to fit together the pieces across horizons
like $r = 2m$.

One can, of course, give criteria that establish {\it non\/}-analyticity,
for example the divergence of invariants formed from the Riemann tensor
and/or its derivatives. This happens, for example, at $r = 0$ in the
Schwarzschild metric, so no real analytic extension is possible there.
It is worth noting that, in the case of indefinite
metrics, not all divergences of the Riemann tensor can be found in this
way; this happens when there is a ``null" infinity, as in a Riemann tensor
of the type $R_{\mu \nu \alpha \beta} = l_{[\mu}m_{\nu]}l_{[\alpha}m_{\beta]}$,
with $l^\mu l_\mu = 0$, $l^\mu m_\mu = 0$ and divergence in $l,\/m$. In
that case the components in an orthonormal frame diverge. (To avoid spurious
infinities due to the frame becoming null the orthonormal frame should be
parallelly propagated).

If one admits an excursion to complex coordinate values one may find
other real metrics analytically related to the original one. Such extensions
are however not unique, and may have nothing directly to do with the
original geometry.\footnote{\small To restore uniqueness it has been
suggested \cite{PSH} that the slightly complex path should be a
geodesic. It remains to be seen whether one does not still lose
physical significance in this unorthodox continuation.}
In fact, the ``extension" may have a different signature than the
original metric. An example is the Euclidean Schwarzschild geometry
that is used to describe instanton or thermal effects. An example of
a Lorentzian, complex analytic but hardly physical relation of the
Schwarzschild metric is
$$ds^2 = \left(1-{2m\over r}\right)dt^2 +
{dr^2\over{\left(1-{2m\over r}\right)}} - r^2 d\theta^2 +
r^2 \cosh^2\!\theta\, d\phi^2.$$
Another example is the continuation of the de Sitter space metric,
$$ds^2 = -{dt^2 \over t^2} + e^{2Ht}(dx^2 + dy^2 + dz^2)$$
across $t = 0$, which effectively
changes the cosmological expansion parameter $H$ into its negative.

As remarked above, a null surface is a natural analyticity boundary, because
``new" information can propagate along such surfaces. On the other hand,
analyticity of a region would be expected to extend to the domain of dependence
of that region. The features of simplicity that allow exact solutions
may have a similar extent, so that the coordinates in which a metric is
originally found tend to be analytic only in such domains, even when the
geometry itself can be extended. Therefore an approach worth trying is
to introduce null coordinates in which the boundary is one of the coordinate
surfaces. The following class of metrics provides an example.

\bigskip \bigskip \noindent
{\large\bf 3 Walker's Spacetimes and their Maximal Extension}

\bigskip \noindent
Walker \cite{WA} considers spherically symmetric ``static" metrics of the form
$$ds^2 = -F dt^2 + {dr^2 \over F} + r^2 d\Omega^2$$
where $F = F(r)$ is the norm of the Killing vector $\partial/\partial t$.
Here this Killing vector is not necessarily timelike (hence the quotes
around ``static") because we allow $F$ to be positive or negative.
$F$ may be an analytic function of $r$, satisfying the Einstein equation,
and range over positive and negative values,
but the metric is clearly non-analytic at the zeros of $F$; the
problem is to find the continuation across these zeros.
(Infinities of $F$ imply infinities of the Riemann tensor, so no analytic
continuation is possible there.) Because the angular part is regular
for $r > 0$, it suffices to confine attention to the two-dimensional
$r,\,t$ part of the metric.

By ``factoring" this two-dimensional part into two integrable null
differential forms,
\begin{equation}
 du = dt + {dr \over F}\/, \qquad dv = dt - {dr \over F}
\end{equation}
we can give the metric the double null form, $ds^2 = -F(u\!-\!v)\,dudv$,
but as a metric this is still singular at $F = 0$. If instead, following
Finkelstein's trick, we use $r$ and only one null coordinate, say $u$,
the metric assumes the nonsingular form
$$ds^2= -F(r) du^2 + 2 du dr,$$
which is analytic wherever $F$ is analytic\footnote{\small In this sentence
the word ``analytic" could be validly replaced everywhere by ``smooth"
or ``$C^n$". In those cases the extension would, of course, not be
unique.} as a function. This metric, then, provides the overlap
necessary to connect two regions with opposite signs of $F$.
This analytic connection between two neighboring regions is illustrated
in the conformal diagram shown in Fig.\ 1.
Here we have assumed that the region $r \rightarrow \infty$ has the usual
asymptotically flat structure, and that there is another zero of $F$
at finite $r$ below $r=a$ (corresponding to the ``roof'' of the figure).
If these structures are different, the blocks may not have a diamond shape,
but the region around the zero, $r=a$, will look the same.

\unitlength=.70mm
\thicklines
\begin{picture}(120.00,100.00)(0,20)
\multiput(0,0)(-30,30){3}{\put(90.00,30.00){\line(1,1){30.00}}}  
\multiput(0,0)(-30,-30){2}{\put(120.00,60.00){\line(-1,1){60.00}}} 
\multiput(0,0)(-30,30){2}{
\bezier{288}(60.00,60.00)(90.00,40.00)(120.00,60.00)} 
\put(110.00,50.00){\line(-1,1){60.00}}  
\bezier{288}(90.00,90.00)(70.00,60.00)(90.00,30.00)  
\bezier{288}(60.00,60.00)(90.00,80.00)(120.00,60.00)  
\thinlines
\put(112.33,74.67){\vector(-4,-3){8.33}}
\put(112.33,74.67){\vector(0,-1){19}}
\put(122.0,76.00){\makebox(0,0)[cc]{$t=$ const}}
\put(90.33,22){\makebox(0,0)[cc]{$t = -\infty$}}
\put(41.00,70.33){\vector(2,1){18.70}}
\put(45.00,63.33){\vector(4,1){23.70}}
\put(45.00,63.33){\vector(1,1){5.60}}
\put(49.67,55.67){\vector(1,0){30.00}}
\put(27,70){\makebox(0,0)[cc]{$r=$ const $<a$}}
\put(30.00,62){\makebox(0,0)[cc]{$r=a$, $F=0$}}
\put(35.5,55.50){\makebox(0,0)[cc]{$r=$ const $>a$}}
\put(93.00,99.33){\vector(-3,-1){24}}
\put(103,101.00){\makebox(0,0)[cc]{$u =$ const}}
\put(103.67,84.67){\vector(0,-1){7.30}}
\put(103.67,84.67){\vector(-1,0){18.67}}
\put(113.5,85.50){\makebox(0,0)[cc]{$t=+\infty$}}
\put(90.00,60.00){\makebox(0,0)[cc]{$F>0$}}
\put(60.00,90.00){\makebox(0,0)[cc]{$F<0$}}
\put(84.67,24.00){\vector(-1,4){3.67}}
\put(96.00,24.00){\vector(1,4){3.67}}
\end{picture}

\bigskip

{\noindent \small
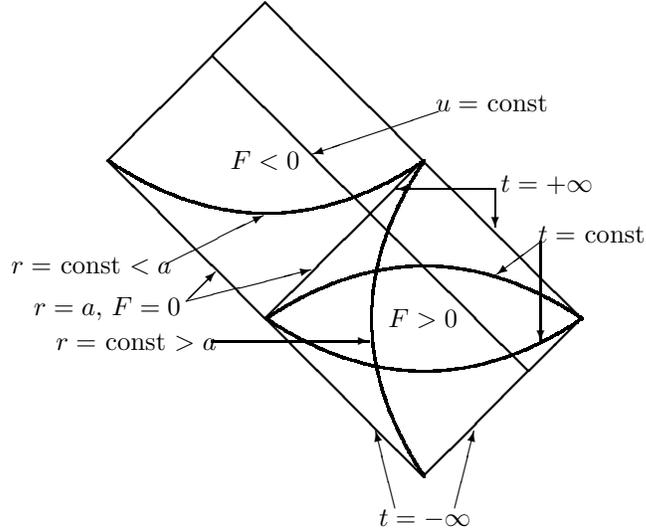
{\bf Fig.\ 1.} Conformal diagram of region of Walker metric
surrounding $r=a$, the largest zero of $F$. The $r,t$
coordinates are degenerate on the boundaries of the diamond-shaped
regions. For example, the lower left boundary can also be labeled
$r=a$.}

\bigskip\bigskip
The spacetime as shown is still not complete. For example, the middle left
boundary $r=a$ is at a finite distance, and so is the ``roof." By using
Finkelstein coordinates $r$ and $v$ we obtain a system that overlaps the
middle left boundary, and by repeating this procedure around the next
smaller zero below $r=a$ we can extend beyond the ``roof." Thus in passing
through each zero of $F$ we add several new diamond-shaped regions
to the conformal diagram, according as we cross the boundary along $u =$ const
or $v =$ const. All the regions so generated at the zero $r=a$ are shown
in Fig.\ 2.

The overlapping coordinates constructed so far reach across all of the
diagonal lines, but not across the intersection points P, Q, R, \ldots.
These come in two types, those like P and R being characterized by the
vanishing of the $\partial/\partial t$ Killing vector, and those like
Q by different values of $r$ trying to come together. It is therefore
not surprising that no analytic continuation is possible
or necessary across points
of type Q; they are at an infinite distance ($t \rightarrow \infty$)
along Killing orbits and not part of the manifold. (They can also provide
a ``safe haven" for observers who might otherwise experience a naked
singularity.) At points of type P the blocks may fit together smoothly
or analytically, depending on the form of $F$. The proof is not immediate;
Walker \cite{WA} introduces lightlike coordinates $U$, $V$ related to
the $u$, $v$ of (1) via an adjustable constant $c$,
$$dU/U = c du \qquad dV/V = -c dv.$$
The metric then takes the form
$$ds^2 = G\, dU dV \qquad {\rm with} \qquad G =
{F\over c^2}\,\exp\left(-2c \int {dr\over F}\right).$$
For functions $F$ of the type
\begin{equation}
F(r) = {\prod_i (r-a_i)\over K(r)}\, ,
\end{equation}
with $K(r)$ a
polynomial with zeros differing from the $a_i$, he shows that $c$ can be
chosen so that $G \neq 0$ at any one $a_i$.

\unitlength=.60mm
\begin{picture}(130.00,141.00)(-15,20)
\thicklines
\put(70.00,20.00){\line(1,1){60.00}}
\put(130.00,80.00){\line(-1,1){60.00}}
\put(70.00,140.00){\line(-1,-1){60.00}}
\put(10.00,80.00){\line(1,-1){60.00}}
\put(40.00,50.00){\line(1,1){60.00}}
\put(100.00,50.00){\line(-1,1){60.00}}
\thinlines
\put(70.00,81.00){\makebox(0,0)[cb]{P}}
\put(101.00,111.00){\makebox(0,0)[cb]{Q}}
\put(70.00,141.00){\makebox(0,0)[cb]{R}}
\put(116.00,96.00){\makebox(0,0)[lb]{$r=\infty,\,t=+\infty$}}
\put(118.00,66.00){\makebox(0,0)[lt]{$r=\infty,\,t=-\infty$}}
\put(82.00,30.00){\makebox(0,0)[lt]{$r=b,\,t=-\infty$}}
\put(94.00,40.00){\makebox(0,0)[lt]{$r=a,\,t=-\infty$}}
\put(94.00,40.00){\vector(-1,3){7.33}}
\put(112.00,106.00){\makebox(0,0)[lc]{$r=a,\,t=+\infty$}}
\put(112.00,106.00){\vector(-3,-1){22.00}}
\put(90.00,124.00){\makebox(0,0)[lb]{$r=b,\,t=+\infty$}}
\put(50.00,122.00){\makebox(0,0)[rb]{$r=b,\,t=-\infty$}}
\put(24.00,96.00){\makebox(0,0)[rb]{$r=\infty,\,t=-\infty$}}
\put(24.00,64.00){\makebox(0,0)[rt]{$r=\infty,\,t=\infty$}}
\put(52.00,34.00){\makebox(0,0)[rt]{$r=b,\,t=+\infty$}}
\put(100.00,80.00){\makebox(0,0)[cc]{$F>0$}}
\put(70.00,50.00){\makebox(0,0)[cc]{$F<0$}}
\put(40.00,80.00){\makebox(0,0)[cc]{$F>0$}}
\put(70.00,110.00){\makebox(0,0)[cc]{$F<0$}}
\end{picture}

\bigskip
{\noindent \small
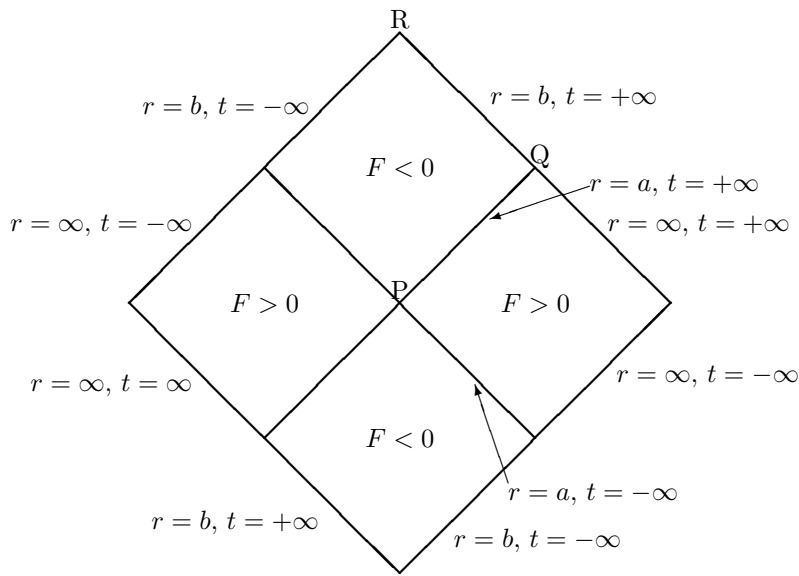
{\bf Fig.\ 2.} Blocks that fit together at their $r=a$ boundary. The
next lower zero of $F$ occurs at $r=b$.}

\bigskip\bigskip

If two roots coincide, the picture looks different.
There are no points of type P because the double-root horizon is
at an infinite spatial distance; a spacelike section does not have
the ``wormhole'' shape, but is an infinite funnel or ``cornucopion.''
Figure 3 shows how the blocks fit together in that case \cite{KL}.
All lines and curves that are shown correspond to $r =$ const.
This diagram looks like what one would obtain by continuing Fig.\ 1
towards the upper left by the usual rules to another block with $F>0$,
and then eliminating the $F<0$ block and moving the two $F>0$ blocks
together. Thus the coincidence limit of two roots of $F$ does not
appear continuous in the conformal picture. This happens, for example,
for the Reissner-Nordstr\"om geometry, where the coincidence limit
corresponds to an ``extremally'' charged black hole, $Q^2 \rightarrow
M^2$. As long as the roots are
distinct the two $F>0$ blocks have a finite size $F<0$ block between
them. Instead of eliminating the $F<0$ block, one can keep its physical size
constant by rescaling the metric. The resulting spacetime is the
Bertotti-Robinson universe. (More detail on the relation between the
extremal Reissner-Nordstr\"om and the Bertotti-Robinson geometries is
found in \cite{CABR}.)

\unitlength=.50mm
\thicklines
\begin{picture}(120.00,120.00)(-30,20)
\multiput(0,0)(-60,60){2}{\put(90.00,30.00){\line(1,1){30.00}}}  
\put(120.00,60.00){\line(-1,1){60.00}} 
\bezier{288}(60.00,120.00)(80.00,90.00)(60.00,60.00)  
\bezier{288}(90.00,90.00)(70.00,60.00)(90.00,30.00)  
\thinlines
\multiput(0,0)(.5,.5){2}{
\put(89.50,29.50){\line(-1,1){60.00}}}
\multiput(0,0)(-.5,.5){2}{\put(60.50,59.50){\line(1,1){30.00}}}
\put(95.00,100.00){\vector(-1,0){25.00}}
\put(96.00,100.00){\makebox(0,0)[lc]{$r =$ const $<a$}}
\put(120.00,75.00){\vector(-1,0){37.00}}
\put(122.00,75.00){\makebox(0,0)[lc]{$r=$ const $>a$}}
\end{picture}

{\noindent \small
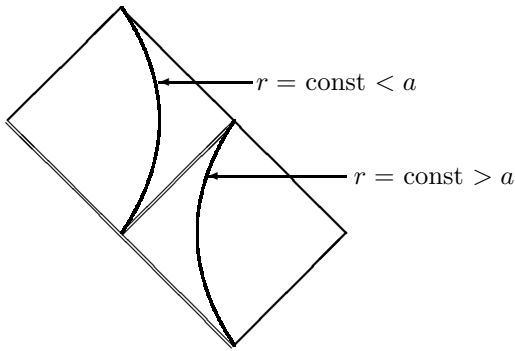
{\bf Fig.\ 3.} Conformal diagram of region of Walker metric surrounding
a double root of $F$, denoted by double lines.}

\bigskip\bigskip
Beyond the coincidence limit it can happen that a pair of (real) roots
disappear. This is also not a continuous change in the conformal
diagram --- the two regions of Fig.\ 3 merge into one.

Metrics of the Walker type occur in the cosmological black hole context
when there is a ``single'' black hole of mass $M$ and charge $Q$ in a
universe with cosmological constant $\Lambda$. In that case, $F$ takes the
form
\begin{equation}
F(r) = {\left(- {\textstyle {1 \over 3}}\Lambda  r^4 + r^2 -
2Mr + Q^2\right)\over r^2}\, ,
\end{equation}
which is of the type (2). (As there, we will denote the zeros of the
numerator, in decreasing order, by $a_1$ \ldots $a_4$.) Thus we know that
the maximal analytic extension is given by the Walker construction, and
once we know how the blocks fit together we can do all calculations in
the original $r$, $t$ coordinates.

\bigskip \bigskip

{\large\bf\noindent
4 Global Structure of de Sitter and Reissner-Nordstr\"om-

\hspace{-3truemm} de Sitter Cosmos}

\bigskip \noindent
{}From the above block-gluing rules it is clear that the conformal diagrams
for the Reissner-Nordstr\"om-de Sitter metrics depend only on the
number of zeros of the function $F$ of (3). Only three of the roots
are positive (for positive $M$ and $\Lambda$). The simplest example is
de Sitter space itself, $M = 0$, $Q = 0$, so $a_1 = \sqrt{3/\Lambda}$,
$a_2=a_3=0$. The blocks look different in this case, because $r=0$ is
a regular origin. Also, $r=\infty$ is infinite distance in time (since
$F < 0$ for $r > a_1$), so we can identify it with timelike and null
infinity, $\Im$. Figure 4a shows a an embedding of an $r,\,t$ subspace of
this geometry in flat 3-dimensional Minkowski space, and Fig.\ 4b is the
corresponding conformal diagram. Note that the conformal diagram corresponds
to only half of the embedded surface, because the latter shows both ``sides''
of the origin ($\phi=0$ and $\phi=\pi$, for example).

\unitlength=0.80mm
\begin{picture}(125.00,73.00)(-5,0)
\thinlines
\put(10.00,10.00){\line(1,1){40.00}}
\put(10.00,50.00){\line(1,-1){40.00}}
\bezier{100}(42.00,32.00)(30.00,28.00)(18.00,32.00)
\thicklines
\bezier{220}(52.00,11.00)(33.00,30.00)(52.00,51.00)
\bezier{228}(8.00,51.00)(28.00,30.00)(8.00,11.00)
\bezier{196}(8.00,11.00)(30.00,0.00)(52.00,11.00)
\thinlines
\put(36.00,47.00){\vector(1,4){2.67}}
\put(39.00,60.00){\makebox(0,0)[cb]{$r \rightarrow \infty\,\,(\Im^+)$}}
\put(52.00,30.00){\vector(-2,-1){9.00}}
\put(53.00,31.00){\makebox(0,0)[lb]{$r=0$}}
\put(51.00,28.00){\makebox(0,0)[lc]{(antipodal}}
\put(52.00,26.00){\makebox(0,0)[lt]{observer)}}
\put(7.00,43.00){\vector(1,0){10.00}}
\put(6.00,43.00){\makebox(0,0)[rc]{$r=a_1$}}
\put(5.00,30.75){\makebox(0,0)[rc]{$t=0$}}
\put(6.00,22.00){\vector(1,0){10.00}}
\put(5.00,22.00){\makebox(0,0)[rc]{$r=0$}}
\put(8.00,18.00){\makebox(0,0)[rc]{(observer)}}
\put(6.00,30.50){\vector(1,0){15.00}}
\put(85.00,10.00){\line(1,1){40.00}}
\put(85.00,10.00){\framebox(40.00,40.00)[cc]{}}
\put(125.00,10.00){\line(-1,1){40.00}}
\put(105.00,51.00){\makebox(0,0)[cb]{$r=\infty \enspace \Im^+$}}
\put(96.00,28.90){\makebox(0,0)[cb]{$t=0$}}
\put(84.00,32.00){\makebox(0,0)[rc]{$r=0$}}
\put(126.00,32.00){\makebox(0,0)[lc]{$r=0$}}
\put(95.00,40.00){\makebox(0,0)[cc]{$r=a_1$}}
\put(115.00,40.00){\makebox(0,0)[cc]{$r=a_1$}}
\put(105.00,10.00){\makebox(0,0)[ct]{$r=\infty \enspace \Im^-$}}
\multiput(0,0)(5,0){8}{\put(86.00,30.00){\line(1,0){3.00}}}
\put(30.00,0.00){\makebox(0,0)[cb]{a}}
\put(105.00,0.00){\makebox(0,0)[cb]{b}}
\thicklines
\bezier{32}(9.83,50.17)(5.83,51.17)(9.83,52.17)
\bezier{168}(9.83,50.17)(30.00,43.50)(50.00,50.00)
\bezier{164}(9.83,52.17)(30.17,56.50)(50.17,52.17)
\bezier{28}(50.00,50.00)(53.67,51.33)(50.17,52.17)
\end{picture}

\bigskip
{\noindent \small
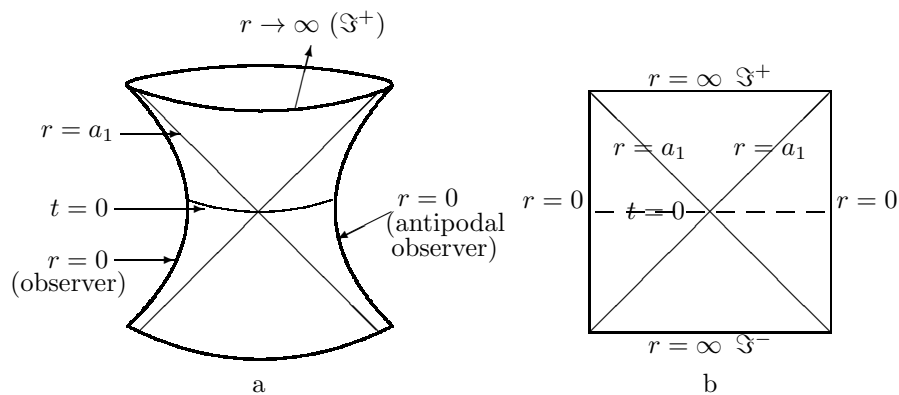
{\bf Fig.\ 4a.} Embedding of 2-D de Sitter space in 3-D
Minkowski space.

\noindent {\bf Fig.\ 4b.} The corresponding conformal diagram.}

\bigskip\bigskip
De Sitter space is often described in coordinates different from the
$r,\,t$ used above, in which the space sections are conformally flat.
These new coordinates $r',\, t'$ are called {\em isotropic} or
{\em cosmological} coordinates \cite{BH},
$$r = r'e^{Ht'} \qquad t = t' - {1\over 2H}\ln (1- H^2r^2).$$
Here $H = \sqrt{\Lambda/3} = 1/a_1$ is the ``Hubble constant" or
cosmological expansion parameter. The metric then becomes
\begin{equation}
ds^2 = -dt'^2 + e^{2Ht'}\left(dr^2 + r^2 d\Omega^2\right).
\end{equation}
These coordinates cover more of de Sitter space than one patch of the
$r,\,t$ coordinates, but there still is a horizon at $t' = -\infty$.
Another block of $r',\,t'$ coordinates  but with opposite sign of
$H$ can be analytically connected to the original one.  We call the
coordinates ``expanding'' in the block with $H > 0$, and ``contracting''
in the other one. Figure 5 shows some of the spacelike surfaces $t'=$ const.

That the de Sitter universe can appear as either static (3), or expanding
or contracting (4), is an accident due to the high degree of symmetry
of this model. In fact, each ``observer" (timelike geodesic)
has a static frame centered around him/her. When there is a ``single''
black hole present, a static frame still exists (but only the one that
is centered about the black hole). The analytic extension of this case
can therefore be easily treated by Walker's method. In discussing the
global properties it is appropriate also to show the expanding and
contracting, cosmological frames, because only their analog exists in
the spacetimes \cite{KT} with several black holes.

\unitlength=.70mm
\begin{picture}(105.00,85.00)(-10,5)
\thinlines
\put(45.00,10.00){\framebox(60.00,60.00)[cc]{}}
\bezier{256}(45.00,10.00)(75.00,25.00)(105.00,25.00)
\bezier{300}(45.00,10.00)(90.00,50.00)(105.00,50.00)
\thicklines
\put(45.00,10.00){\line(0,1){60.00}}
\put(45.00,70.00){\line(1,0){60.00}}
\put(45.00,10.00){\line(1,1){60.00}}
\bezier{300}(45.00,30.00)(60.00,30.00)(105.00,70.00)
\bezier{256}(105.00,70.00)(75.00,55.00)(45.00,55.00)
\thinlines
\put(65.00,50.00){\makebox(0,0)[cc]{$H>0$}}
\put(85.00,30.00){\makebox(0,0)[cc]{$H<0$}}
\put(75.00,71.00){\makebox(0,0)[cb]{$t'=\infty\enspace\Im^+$}}
\put(75.00,10.00){\makebox(0,0)[ct]{$t'=\infty\enspace\Im^-$}}
\put(40.00,25.00){\vector(4,1){26.00}}
\put(38.00,25.00){\makebox(0,0)[rc]{$t'= -\infty$}}
\put(40.00,43.00){\vector(4,-1){23.00}}
\put(38.00,43.00){\makebox(0,0)[rc]{$t'=$ const}}
\end{picture}

\bigskip
{\noindent \small
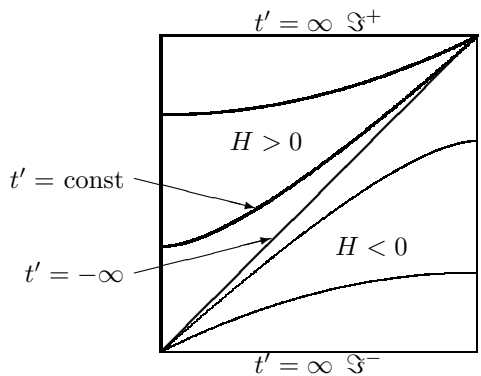
{\bf Fig.\ 5.} Cosmological coordinates in de Sitter space. The part drawn
in thick lines is the expanding region.}

\bigskip\bigskip

The static spherically symmetric solution of the Einstein-Maxwell equations
with a cosmological constant is the Reissner-Nordstr\"om-de Sitter metric
and EM potential (abbreviated RNdS),
$$ds^2 = -F\,dT^2 + {dR^2\over F} + R^2\,d\Omega^2$$
$$F = 1-{2m\over R}+{Q^2\over R^2}-{1\over 3}\Lambda R^2 \qquad
A_T = -{Q\over R}\,.$$
Here we have used capital letters to denote the static frame in order
to distinguish it from the cosmological coordinates, which will be in
lower case.  By means of a somewhat involved coordinate transformation
\cite{BH} one finds the form of the metric in cosmological coordinates,
\begin{equation}
ds^2 = V^{-2} dt^2 + U^2 e^{2Ht} (dr^2 + r^2 d\Omega^2)\,,
\end{equation}
where
$$U = 1+{M\over\rho}+{M^2-Q^2\over 4\rho^2} \qquad
V={U\over 1-{M^2-Q^2\over 4\rho^2}} \qquad \rho=e^{Ht} r\,.$$
By means of the simple coordinate change
$$\tau \equiv H^{-1}e^{Ht}$$
we can extend the region covered by these coordinates, by
allowing $\tau$ to be negative as well as positive. The metric
and EM potential then become
\begin{equation}
ds^2=-{d\tau^2\over U^2}+U^2(dr^2+r^2d\Omega^2) \qquad
U = H\tau+{M\over r} \qquad A_\tau={1\over U}\,.
\end{equation}
Figure 6 shows the analytic extension constructed according to the
prescription of Sect.\ 3 for the generic case, when there are three
roots of $F$.

\unitlength=.5mm
\begin{picture}(60.00,145.00)(-30,-4)
\thinlines
\multiput(0,0)(0,40){2}{
    \put(20.00,40.00){\line(1,0){40.00}}}
\put(100.00,40.00){\line(1,0){40.00}}
\multiput(0,0)(40,0){2}{
    \multiput(0,0)(0,-3){13}{
        \put(60.00,120.00){\makebox(0,0)[ct]{*}}}}
\multiput(0,0)(0,-3){13}{
    \put(60.00,40.00){\makebox(0,0)[ct]{*}}}
\multiput(0,0)(0,-3){13}{
    \put(100.00,40.00){\makebox(0,0)[ct]{\bf*}}}
\multiput(0,0)(80,0){2}{
    \put(40.00,39.00){\makebox(0,0)[ct]{$\Im^-$}}
    \put(40.00,81.00){\makebox(0,0)[cb]{$\Im^+$}}}
\multiput(-40,0)(80,0){2}{        
   \multiput(0,0)(2,-2){20}{
        \put(60.00,80.00){\makebox(0,0)[cb]{.}}}}
\multiput(0,0)(0,40){2}{          
   \multiput(0,0)(2,-2){20}{
        \put(60.00,80.00){\makebox(0,0)[cb]{.}}}}
\multiput(0,0)(2,2){40}{          
        \put(20.00,40.00){\makebox(0,0)[cb]{.}}}
\multiput(0,0)(2,2){20}{          
        \put(60.00,0.00){\makebox(0,0)[cb]{.}}}
\thicklines
\put(100.00,80.00){\line(1,0){40.00}}
\put(60.00,40.00){\line(1,1){40.00}}
\put(100.00,40.00){\line(1,1){40.00}}
\put(60.00,40.00){\line(1,-1){40.00}}
\bezier{180}(60.00,40.00)(80.00,50.00)(100.00,40.00)
\bezier{180}(60.00,40.00)(80.00,30.00)(100.00,40.00)
\bezier{92}(90.00,48.00)(80.00,42.00)(70.00,48.00)
\bezier{244}(90.00,48.00)(130.00,80.00)(140.00,80.00)
\bezier{120}(70.00,30.00)(80.00,40.00)(92.00,30.00)
\bezier{40}(92.00,30.00)(96.00,26.00)(100.00,26.00)
\thinlines
\bezier{244}(70.00,48.00)(30.00,80.00)(20.00,80.00)
\bezier{48}(60.00,26.00)(66.00,26.00)(70.00,30.00)

\end{picture}

\smallskip
{\noindent \small
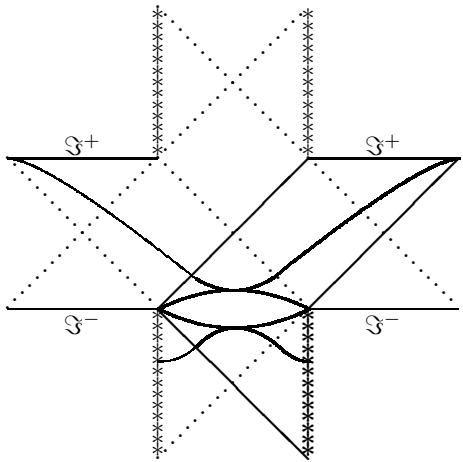
{\bf Fig.\ 6.} Conformal diagram of the RNdS geometry. The diagonal
(mostly dotted) lines are the horizons corresponding to the
three roots of $F$ that separate the different
``static'' blocks. Those crossing between $\Im^-$ and $\Im^+$ are
the cosmological horizons; those crossing at the center of the
figure are the outer black hole horizons; and those crossing at
the top and at the bottom are the inner horizons. The multiply-crossed
vertical lines are the singularities at $R=0$. The thin curves
describe surfaces of constant cosmological time $\tau$, the upper
one having $\tau>0$, and the lower one $\tau<0$. The cosmological
$r$-coordinate is single-valued only to the right (or only to the left)
of the point of contact of these curves with the lens-shaped region.
This point of contact is a ``wormhole throat'' of the spacelike
geometry. Surfaces of different $\tau$-value touch the lens-shaped
region at different points. The boundaries of the lens-shaped region
are given by $R={\rm const}=M\pm\sqrt{M^2-Q^2}$.
The two region covered by the right-hand
parts of these spacelike surfaces, in which the
cosmological coordinates are therefore single-valued, is shown by the
thick outlines. Note that the part near $\Im^+$ is similar to
that shown in Fig.\ 5. The
lens-shaped region in between is not covered by these coordinates. These
regions as drawn are appropriate for expanding coordinates. A region
covered by contracting coordinates is obtained, for example, by
reflecting the regions in thick outline about the horizontal symmetry
axis.}

\bigskip\bigskip
The figure can be repeated indefinitely in the horizontal and the
vertical direction, yielding a spacetime that is spacelike and
timelike periodic. The spacelike surface $T=0$ that cuts the
figure horizontally in its center then has the geometry akin to
a string of beads: as we move along this surface, the radius of
the sphere in the other two ($\theta,\,\phi$) directions alternately
reaches maxima and minima. The maxima correspond to the large
regions of the universe (the ``background de Sitter space''),
and the minima are throats of wormholes that connect one de Sitter
region with the next. Thus each de Sitter region contains two
wormhole mouths, placed in antipodal regions of each large universe.
(This is the reason for the quotes above when calling this a
``single'' black hole in a de Sitter universe.)
Alternatively and more compactly we can imagine the left and right
halves of the figure identified, so that the horizontal spacelike
surface of the figure is a closed circle, and the 3-D spacelike
topology is $S^1 \times S^2$. In this case the electric flux of
the charge $Q$ also describes closed circles: it emerges from one wormhole
mouth, spreads out to the maximum universe size, reconverges on the
other mouth, and flows through the wormhole back to the first mouth.
Seen from the large universe, the first mouth appears positively
charged, and the antipodal one, negatively charged --- an example
of Wheeler's ``charge without charge''!

In this universe (as in the $\Lambda=0$, asymptotically flat
Reissner-Nordstr\"om (RN) geometry) a geodesic observer that
wants to experience the singularity can do so, for example by moving
along the vertical symmetry axis of the figure. In the RN case this
is not considered a serious challenge to cosmic censorship, because
the interior of the black hole, through which the observer in search
of a singular experience must travel, is not stable under small perturbations
of the exterior: radiation falling into the hole from the exterior
would have a large blueshift at this observer --- it would not only
burn her up, but also change the nature of the singularity. It is
remarkable that this does not necessarily happen in RNdS
universes, for certain values of the parameters \cite{MM}.
Thus these solutions
are a counterexample to a strong interpretation of cosmic censorship.
But there are, of course, many other geodesics that can lead observers
who do not take the plunge to their safe haven at $\Im^+$.

The picture of Fig.\ 6 changes, for example as in Fig.\ 3, when roots
of $F$ coincide or cease to be real (see \cite{BH} and \cite{K?}).
Another special case is of interest here because it can be generalized
to the multi-black-hole case, namely  $Q^2 = M^2$.
In contrast to the RN case, when $\Lambda \neq 0$ this choice
does not force a double root, so the global structure and conformal
diagram of Fig.\ 6 still applies. What changes is the way the
cosmological coordinates cover the diagram: the
lens-shaped region degenerates into the line $\tau=0$, so that a
continuous region between the singularity and $\Im^+$ is covered.
Therefore in this case the entire analytic extension of the metric
(6) can be obtained by gluing together alternate copies of expanding
and contracting cosmological coordinate patches. Figure 7 shows a
pair of such regions, with the contracting one outlined by thick lines.
(For easy comparison with Fig.\ 6 it may help to turn the latter
upside down.) We see that the expanding patch contains $\Im^+$,
and the contracting one contains $\Im^-$.
It is therefore not immediately clear,
if we calculate in contracting coordinates only,
how to identify the black hole event horizon as the boundary of the
past of $\Im^+$. We can follow outgoing lightrays to $r=\infty$,
but at that point their geometrical distance $R$, measured by the
Schwarzschild coordinate $R$ (or by the area of the sphere $r=$
const) is still finite, $R = a_1$. However, there is this difference
between such lightrays and those that fall into the black hole, that
the latter reach the geometrical singularity, which {\em is} contained
in the contracting cosmological coordinates. Furthermore, timelike
geodesics heading toward the point S in the figure have an infinite
proper time. Thus S is a safe haven for observers who desire to avoid
the black hole, and it can be regarded as small piece of $\Im^+$ that
can be asymptotically reached in contracting cosmological coordinates,
just enough to be able to identify the event horizon.

\unitlength=.75mm
\thinlines
\begin{picture}(141.00,133.00)(20,0)
\put(60.00,0.00){\line(-1,1){40.00}}
\put(20.00,40.00){\line(1,1){40.00}}
\put(60.00,80.00){\line(1,0){40.00}}
\put(80.00,81.00){\makebox(0,0)[cb]{$\Im^+$}}
\put(80.00,40.00){\makebox(0,0)[ct]{$\Im^-$}}
\thicklines
\put(100.00,80.00){\line(-1,-1){40.00}}
\put(60.00,40.00){\line(1,0){40.00}}
\put(100.00,40.00){\line(1,1){40.00}}
\put(140.00,80.00){\line(-1,1){40.00}}
\bezier{188}(60.00,40.00)(75.00,40.00)(100.00,60.00)
\bezier{188}(100.00,60.00)(125.00,80.00)(140.00,80.00)
\multiput(0,0)(0,-2){20}{
        \put(100.00,120.00){\makebox(0,0)[ct]{\bf*}}}
\thinlines
\bezier{312}(90.00,40.00)(120.00,80.00)(100.00,100.00)
\multiput(0,0)(0,-2){20}{
        \put(60.00,40.00){\makebox(0,0)[ct]{*}}}
\multiput(0,0)(2,-2){20}{
        \put(60.00,80.00){\makebox(0,0)[ct]{.}}}
\multiput(0,0)(2,-2){10}{
        \put(100.00,80.00){\makebox(0,0)[ct]{.}}}
\put(140.00,65.00){\vector(-1,0){25.00}}
\put(141.00,65.00){\makebox(0,0)[lc]{event horizon}}
\put(120.00,43.00){\vector(-1,0){22.00}}
\put(121.00,43.00){\makebox(0,0)[lc]{past de Sitter horizon}}
\put(129.00,110.00){\vector(-4,-1){15.00}}
\put(130.00,110.00){\makebox(0,0)[lc]{$r=0$}}
\put(130.00,99.00){\vector(-4,-1){24.00}}
\put(131.00,99.00){\makebox(0,0)[lc]{$r=$ const}}
\put(50.00,50.00){\makebox(0,0)[cc]{expanding}}
\put(122.00,82.00){\makebox(0,0)[cc]{contracting}}
\put(140.00,73.00){\vector(-3,1){12.00}}
\put(141.00,73.00){\makebox(0,0)[lc]{$\tau=$ const$<0$}}
\put(90.00,90.00){\vector(1,-1){9.00}}
\put(89.00,91.00){\makebox(0,0)[rb]{S}}
\put(46.00,73.00){\vector(2,-1){31.00}}
\put(45.00,73.00){\makebox(0,0)[rb]{$r=\infty$}}
\put(32.00,63.00){\vector(2,-1){7.00}}
\put(31.00,63.00){\makebox(0,0)[rc]{$r=0$}}
\end{picture}

\vspace{.5truecm}
{\noindent \small
{\bf Fig.\ 7.} Two patches of cosmological coordinates for the RNdS
geometry for the case $Q^2=M^2$ and $p<1$ (``undermassive'').}

\bigskip\bigskip
The $Q^2=M^2$ RNdS geometries still depend on two parameters, $M$
and $\Lambda$ resp.\ $H$, or on one dimensionless parameter $p=4M|H|$
up to scale change. If $p<1$ we have the ``undermassive'' case
discussed so far. If $p>1$ there is only one real root of $F(R)=0$.
The outer black hole horizon and the
de Sitter horizon have disappeared, only what used to be
the inner black hole horizon remains. One could also interpret the
remaining horizon as a cosmological one, separating two naked
singularities at antipodal regions of a background de Sitter space.
The conformal diagram for this case, constructed according to the
block gluing rules, is shown in Fig.\ 8.

\unitlength=.75mm
\begin{picture}(106.00,60)(22,0)
\thinlines
\multiput(0,0)(0,-2){20}{
\put(100.00,45.00){\makebox(0,0)[ct]{*}}}
\multiput(0,0)(2,2){20}{
\put(60.00,5.00){\makebox(0,0)[ct]{.}}}
\put(60.00,45.00){\line(1,0){40.00}}
\put(80.00,46.00){\makebox(0,0)[cb]{$\Im^+$}}
\bezier{128}(60.00,30.00)(76.00,16.00)(76.00,5.00)
\put(80.00,4.00){\makebox(0,0)[ct]{$\Im^-$}}
\put(105.00,20.00){\vector(-2,-1){20.00}}
\put(106.00,20.00){\makebox(0,0)[lc]{$\tau=$ const $<0$}}
\put(105.00,30.00){\vector(-2,-1){26.00}}
\put(106.00,30.00){\makebox(0,0)[lc]{$\tau=$ const $>0$}}
\put(50.00,24.00){\vector(1,0){15.00}}
\put(49.00,24.00){\makebox(0,0)[rc]{$r=$ const}}
\put(71.50,13.00){\makebox(0,0)[cc]{\small contracting}}
\put(85.00,35.00){\makebox(0,0)[cc]{\small expanding}}
\thicklines
\put(60.00,5.00){\line(1,0){40.00}}
\put(100.00,5.00){\line(-1,1){40.00}}
\bezier{176}(60.00,5.00)(79.00,14.00)(100.00,5.00)
\bezier{176}(60.00,20.00)(80.00,19.00)(100.00,5.00)
\multiput(0,0)(0,-2){20}{
\put(60.00,45.00){\makebox(0,0)[ct]{\bf*}}}

\end{picture}

\vspace{.6truecm}
{\noindent \small
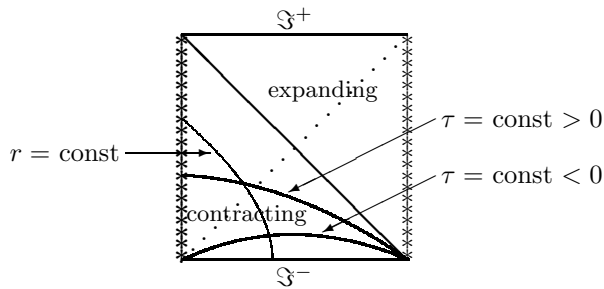
{\bf Fig.\ 8.} Two patches of cosmological coordinates for the RNdS
geometry for the case $Q^2=M^2$ and $p>1$ (``overmassive''). The thick
outline shows the contracting patch.}

\bigskip \bigskip \noindent
{\bf 4.1 Collapsing Dust}

\bigskip \noindent
The solutions discussed so far correspond to ``eternal'' black holes (or a
combination of white holes and black holes). The future part of the usual black
hole geometry can be generated from matter initial data, say by the collapse
of  a sphere of dust. A similar process is possible for $Q^2=M^2$ RNdS.

To generate such a geometry from dust, the dust must itself have $Q^2=M^2$,
so that electrical forces largely balance gravitational forces. The dust can
still
collapse if it has the right initial velocity. A simple situation is dust that
is at
rest in cosmological coordinates: in any metric and potential of form (6), for
an arbitrary $U$, such dust remains at rest ($r=$ const), if considered as
test particles. A typical wordline is shown in Fig.\ 7. To understand why
it behaves this way we look more closely at the electric field associated
with this geometry. We know that the parameter $Q$ of the black hole on
the left is opposite to that of the one on the right. Suppose the left
hole is negative and the right one positive. On the spacelike surface
corresponding to a horizontal line drawn through the center of
Fig.\ 7, the electric field will then point from right to left. By flux
conservation, the electric field will therefore also point to the left
in the region near the upper singularity shown in the Figure. Thus we
can associate a {\em negative} charge with this singularity. (The
singularity to the right of the $r=0$ inner horizon, which is not shown
in the Figure, correspondingly has positive charge). The dust particle
on the $r=$ const trajectory has the same charge as the black hole that
is included in that coordinate patch, namely positive. It therefore ends
up on the negative singularity.

Any point in Fig.\ 7 can be considered to be in either of two possible
cosmological coordinate patches. For example, a point near $\Im^-$ is
in the contracting patch shown, which contracts about the right black
hole. By reflecting this patch about a vertical line through the center
of the Figure we obtain a patch that contracts about the left black hole.
It also contains the region near $\Im^-$. Its radial coordinate will be
denoted by $r'$. The trajectory $r'=$ const, obtained by reflecting the
$r=$ const trajectory shown in the Figure, describes a negative test charge
that falls into the positive singularity of the left (negative) black hole.
Similarly the region between the de Sitter and the event horizon in the
contracting patch has trajectories of positive charges that fall into the
positive black hole, and of negative charges that go to $\Im^+$. The
region inside the event horizon has trajectories that go to one or the
other singularity, depending on their charges. Of course all these
trajectories that are simply described by $r=$ const or $r'=$ const
satisfy special initial conditions --- their initial position is arbitrary,
but their initial velocity is then determined.

To take into account the effect
of the dust {\em on} the metric and potential, one needs to match the vacuum
region to an interior solution. For spherical symmetry the matching conditions
are equivalent to the demand that the dust at the boundary of the interior
region
move on a test particle path of the vacuum region. Thus a possible boundary
for a collapsing (expanding) ball of dust is $r=$ const in collapsing
(expanding) cosmological coordinates. The surface area $4\pi r^2 U^2$ of the
dust ball collapses to zero when $U = 0$, which is also the location of
the geometrical singularity; at that point the center of the dust ball must
coincide with the surface.  The corresponding conformal diagram therefore
looks as shown by the thick curves in Fig. 9. The region filled with dots
denotes the location of the dust.

\unitlength=2.0mm
\begin{picture}(40.00,28.00)(0,5)
\thicklines
\put(10.00,10.00){\line(1,0){13.00}}
\put(20.00,20.00){\line(-1,-1){10.00}}
\bezier{20}(25.00,18.33)(26.50,20.00)(25.00,22.00)
\put(18.20,10.00){\line(5,6){6.92}}
\put(20.00,28.00){\line(5,-6){5.08}}
\bezier{44}(25.00,23.00)(29.00,19.00)(26.00,15.00)
\put(20.00,28.00){\line(1,-1){5.00}}
\put(26.08,15.00){\line(-3,-5){3.00}}
\thinlines
\put(30.00,28.00){\line(-5,-6){5.00}}
\put(32.00,10.00){\line(-5,6){7.08}}
\bezier{20}(25.00,18.33)(23.17,20.00)(25.00,22.00)
\put(40.00,10.00){\line(-1,1){10.00}}
\put(30.00,28.00){\line(-1,-1){5.00}}
\bezier{44}(25.00,23.00)(21.00,19.00)(24.00,15.00)
\put(24.00,15.00){\line(3,-5){3.00}}
\put(27.00,10.00){\line(1,0){13.00}}
\multiput(0,0)(0,0.75){11}{
\put(20.00,20.00){\makebox(0,0)[cc]{\bf *}}}
\multiput(0,0)(.8,-.8){7}{
\put(20.60,19.40){\makebox(0,0)[cb]{\bf.}}}
\multiput(0,0)(0,0.75){11}{
\put(30.00,20.00){\makebox(0,0)[cc]{*}}}
\put(20.00,11.00){\circle*{0.00}}
\put(21.00,11.00){\circle*{0.00}}
\put(22.00,11.00){\circle*{0.00}}
\put(23.00,11.00){\circle*{0.00}}
\put(24.00,12.00){\circle*{0.00}}
\put(23.00,12.00){\circle*{0.00}}
\put(22.00,12.00){\circle*{0.00}}
\put(21.00,12.00){\circle*{0.00}}
\put(21.00,13.00){\circle*{0.00}}
\put(22.00,13.00){\circle*{0.00}}
\put(23.00,13.00){\circle*{0.00}}
\put(24.00,13.00){\circle*{0.00}}
\put(25.00,14.00){\circle*{0.00}}
\put(24.00,14.00){\circle*{0.00}}
\put(23.00,14.00){\circle*{0.00}}
\put(22.00,14.00){\circle*{0.00}}
\put(23.00,15.00){\circle*{0.00}}
\put(25.00,15.00){\circle*{0.00}}
\put(26.00,16.00){\circle*{0.00}}
\put(25.00,16.00){\circle*{0.00}}
\put(24.00,16.00){\circle*{0.00}}
\put(25.00,17.00){\circle*{0.00}}
\put(27.00,18.00){\circle*{0.00}}
\put(26.00,18.00){\circle*{0.00}}
\put(26.00,19.00){\circle*{0.00}}
\put(27.00,19.00){\circle*{0.00}}
\put(26.00,20.00){\circle*{0.00}}
\put(27.00,20.00){\circle*{0.00}}
\put(26.00,21.00){\circle*{0.00}}
\put(25.50,22.00){\circle*{0.00}}
\put(24.92,22.67){\circle*{0.00}}
\put(24.33,23.25){\circle*{0.00}}
\put(23.75,23.92){\circle*{0.00}}
\put(23.17,24.58){\circle*{0.00}}
\put(22.50,25.25){\circle*{0.00}}
\put(26.00,17.00){\circle*{0.00}}
\put(32.00,24.00){\vector(-2,-1){5.50}}
\put(32.25,24.00){\makebox(0,0)[lc]{$r=0$}}
\put(18.00,24.00){\vector(2,-1){5.33}}
\put(17.75,24.00){\makebox(0,0)[rc]{$r'=0$}}
\put(14.00,18.00){\vector(1,-1){2.00}}
\put(14.00,18.00){\makebox(0,0)[rc]{$r=\infty$}}
\put(36.25,18.25){\makebox(0,0)[lc]{$r'=\infty$}}
\put(36.00,18.00){\vector(-1,-1){2.00}}
\end{picture}

\vspace{-.6truecm}
{\noindent \small
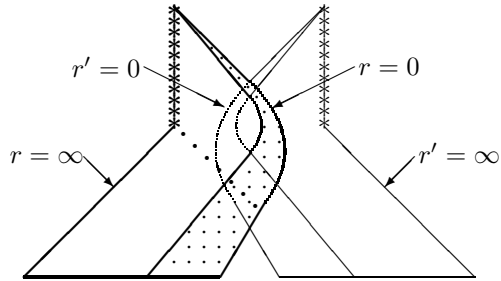
{\bf Fig.\ 9.} Conformal diagram for a RNdS black hole generated by the
collapse of a ball of charged dust in a de Sitter background (thick
lines). The dust region is shown dotted. The dotted line is the black hole
event horizon. The thin lines show a dust ball that is symmetrically
placed at the antipodal region of the RNdS background, and has
opposite charge. If both dust balls are present only the region between
the curves $r=0$ resp.\ $r'=0$ applies.
}

\bigskip\bigskip
Because the dust includes the origin $r=0$, there is now no continuation
to the right of the heavily outlined region necessary or possible. There
is still a cosmological horizon, $r=\infty$, and the continuation on the
other side could be analytic (a semi-infinite ``string of beads'') or
reflection-symmetric about a vertical axis through $\Im^-$ (a collapsing
dust ball at the antipodal region). Of the two geometrical singularities
associated with one RNdS black hole, the dust covers up only the one that
would have been on the right in the figure, with the same total charge as
that of the dust. The other singularity could be covered up by another
dust ball, with the opposite charge. This is shown by the thin curves
in Figure 9. If both dust spheres are present, then the physical part of
the geometry lies between the $r'=0$ curve on the left and the $r=0$ curve
on the right. If conditions are as shown, they uncover and then cover up
again a (small) region of vacuum RNdS geometry between them.

\bigskip \bigskip \noindent
{\large\bf 5 The Multi-Black-Hole Solutions}

\bigskip \noindent
A solution representing any number $n$ of arbitrarily placed charged black
holes in a de Sitter background is given by a metric of type (6), with
a different potential $U$ \cite{KT}:
\begin{equation}
ds^2=-{d\tau^2\over U^2}+U^2(dr^2+r^2d\Omega^2) \qquad
U = H\tau+\sum_{i=1}^n{M_i\over |r-r_i|} \qquad A_\tau={1\over U}\,.
\end{equation}
We will call this the KT solution.  Each mass of the KT solution has
a charge proportional to the mass, $Q_i=M_i$, and only
the location $r_i$ (not the initial velocity) is arbitrary. Here
$|r-r_i|$ denotes the Euclidean distance between the field point $r$
and the fixed location $r_i$ in a Euclidean space of
cosmological coordinates. For $n=1$
this reduces to the $Q^2=M^2$ case of the RNdS solution (6). Also, in the
limit of large $r$, and for $r_i$ in a compact region of Euclidean
coordinate space, (7) approaches the RNdS solution with $M = \sum M_i$.
One therefore expects the horizon structure at $r \rightarrow \infty$ to
be similar to that of RNdS, which suggests that a $H>0$ and a $H<0$ version
of (7) can be glued together as extensions of each other, similar to
Fig.\ 7. The surprise is that, although this can be done with some degree
of smoothness, it cannot be done analytically \cite{BHKT}.
This means that there
is no unique extension. An observer at rest in contracting ($H<0$)
cosmological coordinates, whose entire past can be described in these
coordinates, has no way of telling what is in the other ``half'' of
the de Sitter background  (he can only guess that the total charge over
there must be $-\sum M_i$, to balance the charge that he sees). If he moves
and crosses the cosmological horizon $r=\infty$, the other, expanding half
suddenly comes into view, seen at a very early time when all the masses are
very close together. It is therefore reasonable
that there is a pulse of gravitational and EM radiation associated with
the horizon: this is the physical description of the lack of analyticity.

It is instructive to note how the various solutions differ in respect to
possible coordinate choices. Pure de Sitter space has a static and two
cosmological (expanding and contracting) coordinate systems centered
about any timelike geodesic. In
RNdS these coordinates are centered about the black
hole only, but one can still choose between expanding and contracting
frames near either of the holes. In KT there is one set of black holes,
all with the same sign of charge, that is uniquely expanding (the distances
between holes increasing as $e^{H\tau}$), and another, oppositely
charged set that is uniquely contracting (the distances decreasing as
$e^{-|H|t}$). The expanding set is described by cosmological coordinates
that include $\Im^+$, whereas for the contracting set, $\Im^-$ is
included.

\bigskip \bigskip \noindent
{\bf 5.1 Merging Black Holes}

\bigskip \noindent
Since black (as opposed to white) holes are determined from $\Im^+$, it is
easiest to identify the black hole horizons for the expanding set. Consider
the expanding coordinates in RNdS as in the left half of Fig. 7. The
boundary of the past of $\Im^+$ that lies in those coordinates is the
left-hand line labeled $r=0$ --- it is the event horizon of the black hole
in the ``left'' part of RNdS space. Similarly the event horizon of
expanding KT space is given by $r - r_i = 0$. The Euclidean coordinate
space with the $n$ points $r_i$ removed is a representation of $\Im^+$
of expanding KT space. The $n$ missing points represent $n$ disjoint
boundary components, and there is a finite distance between them at all
times. Thus we have $n$ black holes that remain separate for all times.

It is not as easy to identify the black holes in the contracting KT space,
because that does not contain much of $\Im^+$. But because the cosmological
horizon at $r=\infty$ is so similar to that of the RNdS space it does contain
a ``point'' like S of Fig.\ 7 that is just enough to define the event
horizon. So, to find the event horizon while staying within one coordinate
patch we must find the surface that divides lightrays reaching $r=\infty$
from those that reach the singularity. But where in the KT world
is the singularity? One can show that
metrics of type (7) are singular where $U=0$. In the contracting case,
$H<0$ (and of course $M_i>0$), this happens only for positive $\tau$.
Thus we need to find the ``last" lightrays that just make it out to
$r=\infty$ at $\tau=0$. More precisely (since $r=\infty$ is not a very
precise place) we need $H r\tau$ finite in this limit.\footnote{This
is so because part of the ``somewhat involved" coordinate transformation
leading to (5) is actually rather simple, $R=Hr\tau+M$. For RNdS this
is the static $R$, which is finite on the black hole horizon. Because
near $r=\infty$ the KT geometry is so close to RNdS, $R$ should also
be finite for KT.}

If we know these last lightrays (the 3D horizon surface) we can then
intersect them with a spacelike surface to find the shape of the 2D
horizon at different times. An interesting question is whether this
2D horizon changes topology with time. This would, for example, describe
the merger of two black holes into one. If black holes merge we can try
to make an overmassive (hence nakedly singular) one from two undermassive
ones, to test cosmic censorship. This is of course possible only in a
contracting part of the KT solution, because we have already seen that
the black holes in the expanding part remain separate for all times.

All contracting black holes do eventually merge into one. To show this
for the case of a pair of holes, we show that the horizon must consist
of two parts at early times, and be a single surface at late times. One
can see rather directly that light starting at $r$ sufficiently close to
any one of the $r_i$ at any time and in any direction will reach the
singularity, because it will spend its entire history in a geometry
sufficiently close to the single black hole, RNdS geometry. Thus points
sufficiently close to $r_i$ will always lie within the event horizon.
So, for the case of two black holes, a key question is what happens to
light that starts on the midplane between the holes. In fact, if the light
starts early enough it will always be able to escape to $r=\infty$. Thus
at early times the midplane does not meet the horizon --- the two black
holes are disjoint. To show this we center our Euclidean coordinates
at the midpoint between the holes, which have a Euclidean separation $d$.
{}From $ds^2=0$ in (7) we then find, for radial outgoing null geodesics in the
midplane,
$${d\tau\over dr} = U^2 = \left(H\tau+{M\over\sqrt{r^2+d^2}}\right)^2
< \left(H\tau+{\sqrt{2}M\over {r+d}}\right)^2.$$
Now let $$R_*=H\tau (r+d) \qquad y_*=\ln(r+d)$$ to find
\begin{equation}
(r+d)\, dR_*/dr > R_* +H\,(R_*+\sqrt{2}M)^2.
\end{equation}
Standard analysis of this equation shows that if $R_*$ is larger than
the lower root of the RHS of (8), it will stay positive
for all larger $r$. In terms of $r$ and $\tau$ this means that if $\tau$
is sufficiently negative for any $r$ (remember $H<0$) then $\tau$ will
remain negative as $r$ increases --- the lightray avoids the singularity.
By a similar estimate one can show \cite{BHKT}
that for each sufficiently late
(but negative) $\tau$ there is a sphere surrounding both $r_i$ such
that all outgoing null geodesics will reach the $U=0$ singularity. This
means that the horizon surrounds both $r_i$, and the black holes have
merged.

\bigskip \bigskip
\noindent
{\bf 5.2 Continuing Beyond the Horizons}

\bigskip \noindent
We could now discuss the merging of two undermassive into one overmassive
black hole. Since the geometry near $r=\infty$ will be determined by the
overmassive sum of the two individual masses, that neighborhood will
look like the corresponding part of a single overmassive RNdS, that is,
like the left hand side of Fig.\ 8. That contains a naked singularity
(the left multiply-crossed line), which has nothing to do with the black
hole merger, because it is located at the opposite side of the universe.
But this is the only place in the coordinate patch where a $U=0$ singularity
occurs. Our patch does not describe enough of the history of the interesting
region, where $|r-r_i|$ is small, just as the thick outline does not extend
far to the right side of Fig.\ 8. To find out whether black hole merger
generates its own naked singularity we must continue the KT metric beyond
the inner black hole horizons at $r=r_i$. As we are continuing
the KT geometry it is of course also interesting to look beyond the
cosmological horizon at $r=\infty$, which exists only if the total mass
is undermassive ($4|H|\sum M_i<1$). So we look at null geodesics that
approach these horizons.

Let us choose the origin $r=0$ of our Euclidean coordinates at the location
of the $i^{\rm th}$ mass (so that $r_i=0$). The equation $ds^2=0$ satisfied
by an ingoing null geodesic then takes the form, for small $r$,
\begin{equation}
 {d\tau\over dr}=-U^2=-\left(H\tau+{M\over r}+
\sum_{j\neq i}{M_j\over r_j}\right)^2.
\end{equation}
We can eliminate the last (constant) term on the right by defining a new
time coordinate $\tau'=\tau +H^{-1}\sum'(M_j/r_j)$. The equation then becomes
an equality version of (8), and by analyzing it in the same way as above one
finds \cite{BHKT} the limiting forms
\begin{equation}
2H^2r\tau' \rightarrow 1-2M_iH-\sqrt{1-4M_iH}.
\end{equation}
To assess any incompleteness we need to know how a null geodesic
$(r(s),\,\tau(s))$ depends on the affine parameter $s$, and we can get that
from the variational principle,
$$\delta \int \left(-{1\over U^2}\left({d\tau\over ds}\right)^2 +
U^2\left({dr\over ds}\right)^2\right)ds=0.$$
The Euler-Lagrange equation for $\tau(s)$ together with the first equality
of (9) yields
$${{d^2r}\over{ds^2}}-2HU\left({dr\over ds}\right)^2=0.$$
Substituting (10) we find, in the limit $r \rightarrow 0$,
$${d^2r\over ds^2}-{1-\sqrt{1-4M_iH}\over r}\left({dr\over ds}\right)^2=0$$
with the solution
\begin{equation}
r \sim \left(s-s_{\rm hor}\right)^{1\over\sqrt{1-4M_iH}}, \quad {\rm hence}
\quad \tau\sim \left(s-s_{\rm hor}\right)^{-{1\over\sqrt{1-4M_iH}}}.
\end{equation}
So the inner horizon is reached at a finite parameter value $s_{\rm hor}$.
Similarly one finds that the cosmological horizon is reached in a
finite parameter interval,
\begin{equation}
r \sim\left(s-s_{\rm Hor}\right)^{-{1\over\sqrt{1+4(\Sigma M_i)H}}}, \qquad
\tau \sim \left(s-s_{\rm Hor}\right)^{1\over\sqrt{1+4(\Sigma M_i)H}}.
\end{equation}

This behavior of the coordinates $r$ and $\tau$ gives us important information
about the differentiability and analyticity of the geometry near the horizon.
We can first eliminate whichever of the two is infinite on a given horizon
in favor of $\hat R=Hr\tau$, which is always finite on the horizon.
The metric is then an analytic function of remaining, finite
coordinates, so the Riemann tensor will also be analytic
in these coordinates. But in order that the geometry
be differentiable, the Riemann tensor should be differentiable
in the affine parameter $s$ along null geodesics. Thus the
differentiability of the geometry is measured by that of
$r$ resp.\ $\tau$ as a function of $s$, as given by (11) and (12).

Consider first the neighborhood of the inner horizon, where $r$ is finite.
Since $H<0$ we have $1/\sqrt{1-4M_iH} < 1$, $r$ is not a differentiable
function of $s$ at $r=0$, and the Riemann tensor will be singular there.
A more careful analysis \cite{BHKT}, using a transformation to coordinates
that are not singular on the horizon, shows that the metric is $C^1$ but
in general not $C^2$. There is therefore no unique, analytic extension across
the inner horizon. One can match differentiably essentially any KT solution
with the same mass $M_i$.
One can increase the differentiability by arranging
the other masses carefully around the $i^{\rm th}$ one, so as to make
the potential $U$ approximately spherically symmetric (by eliminating
multipoles to some order). The neighborhood of $M_i$ then becomes
approximately RNdS and hence ``more nearly analytic'' --- i.e., of
increased differentiability.

The situation near the cosmological horizon offers more variety. To have this
horizon at all the total mass must be undermassive, $4|H|\Sigma(M_i)<1$.
Here $\tau$ is the finite one, and the corresponding power of $s$ is
$1/\sqrt{1-4(\Sigma M_i)|H|}>1$. Thus $\tau$ is always at least $C^1$.
The transformation
to coordinates that are good on the horizon shows that the metric is
always at least $C^2$. In the special
cases when the power is an integer $n$, i.e., for masses such that
$$4H\sum M_i = 1 - {1\over n^2},$$
the metric is $C^\infty$. For these values the smooth continuation matches
the KT spacetime at the cosmological horizon to one with the same
position and magnitudes of all the masses (so that all multipole moments
agree), but with the opposite sign of $H$. We do not understand
the physical significance of these special masses.

To show that the geometry at these horizons is not more differentiable
than claimed one can compute the Riemann tensor. This infinity is of the
null type mentioned in Sect.\ 2, and does not show up in invariants
formed from the Riemann tensor. To see this for the horizon at $r=0$
(or $r=\infty$) we write the metric (7) in terms of the coordinate
$\hat R=Hr\tau$ and $y=\ln r$, and the quantity $W=rU$. Then the horizon occurs
at $r \rightarrow \pm \infty$, where $\hat R$ and $W$ are finite,
$$ds^2 = -{(d\hat R-\hat Rdy)^2\over H^2W^2}+W^2(dy^2+d\Omega^2).$$
Now an invariant formed from the curvature tensor involves terms
in derivatives of the metric and its inverse, multiplied by powers
of the metric and its inverse. All these reduce to derivatives of $W$
and $\hat R$ divided by powers of $W$. But all derivatives of $W$ remain
bounded
as $y\rightarrow \pm\infty$, and $W$ is finite on the horizon. Thus the
invariants cannot blow up.

Singularities do show up in the components of the Riemann tensor in a
parallelly propagated frame, for example along the null geodesic
$(r(s),\,\tau(s))$ discussed above. Let $l=\partial/\partial s$ be
the parallelly propagated tangent. Because of the asymptotic symmetry
near the horizon, $\eta = \partial/\partial\theta$ is also asymptotically
parallelly propagated. Now the frame component
$R_{\mu\nu\rho\sigma}l^\mu\eta^\nu l^\rho\eta^\sigma$ contains the term
$g_{\theta\theta,ss}$. If $g_{\theta\theta}=W^2$ depends on $r$ (and not just
on the regular $R$), it will not be a smooth function of $s$. In the
RNdS (``single mass'') case, $g_{\theta\theta}$ depends only on $R$.
In the KT case the corrections to that behavior near an inner ($r=0$) horizon
start with the power $r^2 = \left(s-s_{\rm hor}\right)^{2\over\sqrt{1-4M_iH}}
$ unless there is special symmetry; thus one finds the differentiability
of the metric as claimed above.

\bigskip\bigskip

{\large\bf\noindent
6 Naked Singularities?}

\bigskip \noindent
Naked singularities visible to observers safely outside the strong
curvature regions do not form in realistic gravitational collapse ---
this is the essential notion behind cosmic censorship. It can be made more
precise in various ways \cite{xx}. At present none of these have been proved
to be true in generic cases. When a proof seems difficult, it may be
easier to obtain a convincing counterexample. Even if the conjecture is correct
under certain assumptions, counterexamples are useful
to test the necessity of these assumptions. For example, it may be that a
version of cosmic censorship holds in pure general relativity, but
fails when the theory is modified, say by a cosmological constant or
by applying it to higher dimensions. There are in fact indications that
cosmic censorship fails in the higher-dimensional theory inspired by
string theory:  certain 5D black strings (objects that appear
four-dimensionally like black holes) are unstable, it is entropically
favorable for them to decay into a set of 5D black holes, and
during this decay naked singularities would form \cite{GF}.
In the present contribution we want to test whether cosmic censorship
fails in the special circumstances that are afforded when a cosmological
constant is present.\footnote{A similar test for Einstein-Maxwell-dilaton
theory with a cosmological constant inspired by string theory has been
discussed in \cite{HH}}

The idea of using KT spacetimes to test cosmic censorship is to start
with two or more small (and hence not naked) black holes and
let them collapse to form a single large (and maybe naked) one.
We have seen that the KT solutions indeed can describe coalescing
black holes, in the sense that the event horizons coalesce. But if
it is possible to define the event horizon as we did, by observers who
live for an infinite proper time in one KT coordinate patch, then these
observers will see no signal from any singularity
--- all singularities that form from
the initial data, including those in any of the (non-analytic) extensions,
lie inside this event horizon. To find a situation where there is no
``safe haven,'' so that the generic observers does see a singularity,
we must suppose that $\sum M_i$ is overmassive, so the initial black
holes cannot be defined by their event horizon. An alternative to
starting with black holes that have event horizons is to start with regular
initial data on a compact surface. The KT solution cannot provide
this either, because each $M_i$ has an infinite throat. But such throats
are the next best thing: each throat is undermassive, it is surrounded
by a trapped surface,
so one would not expect that the asymptotic regions down the throat
could influence the solution in the interior. Can we, then, construct
a KT solution of undermassive throats that has regular initial data
and a naked singularity in its time development?

To decide this we must explore the global structure of the KT geometry.
Unfortunately this cannot be completely represented by 2D conformal
diagrams, because there is insufficient symmetry to suppress the
additional dimensions. If we confine attention to the case of two
equal masses, they lie on a line in the Euclidean coordinate
space (which is an axis of symmetry of the spacetime). We can represent
the essential features of the spacetime by drawing the conformal
diagram for the spacetime spanned by the part of the axis going from
one of the masses to infinity. This is shown in Fig.\ 10a.
The part of the diagram
representing the region near $r = \infty$ is to be read like
a normal conformal diagram (i.e., each point represents a 2-sphere),
whereas the region near $r=0$ is to be thought of as doubled (each
point represents two 2-spheres).

\unitlength=1.30mm
\begin{picture}(50.00,53.50)(20,4)
\thinlines
\put(10,0){
\multiput(0,0)(0,1.2){25}{
\put(20.00,10.00){\makebox(0,0)[cc]{{\bf*}}}}
\multiput(20,10)(0,1.2){25}{
\put(20.00,10.00){\makebox(0,0)[cc]{*}}}
\put(20.00,40.00){\line(1,1){10.00}}
\put(30.00,50.00){\line(1,0){10.00}}
\put(15.00,25.00){\vector(1,0){4.50}}
\put(14.00,25.00){\makebox(0,0)[rc]{$r'=0$}}
\put(30.00,20.00){\makebox(0,0)[rc]{$r=0$}}
\put(31.00,21.00){\vector(1,1){4.00}}
\put(31.00,19.00){\vector(1,-1){4.00}}
\put(25.00,10.00){\makebox(0,0)[ct]{$\Im^-$}}
\put(35.00,50.00){\makebox(0,0)[cb]{$\Im^+$}}
\thicklines
\put(20.00,10.00){\line(1,0){10.00}}
\put(30.00,10.00){\line(1,1){10.00}}
\put(40.00,20.00){\line(-1,1){20.00}}
}
\thinlines
\put(80.00,10.00){\line(1,0){10.00}}
\put(90.00,10.00){\line(1,1){10.00}}
\put(85.00,10.00){\makebox(0,0)[ct]{$\Im^-$}}
\put(82.00,24.00){\vector(1,0){4.00}}
\put(81.00,24.00){\makebox(0,0)[rc]{$r'=0$}}
\put(102.00,21.00){\makebox(0,0)[lc]{$r=0$}}
\put(102.00,20.00){\vector(-3,-2){6.00}}
\put(102.00,22.00){\vector(-3,1){6.00}}
\put(82.00,11.00){\circle*{0.00}}
\put(84.00,11.00){\circle*{0.00}}
\put(82.00,13.00){\circle*{0.00}}
\put(84.00,13.00){\circle*{0.00}}
\put(82.00,15.00){\circle*{0.00}}
\put(84.00,15.00){\circle*{0.00}}
\put(88.00,25.00){\circle*{0.00}}
\multiput(80,10)(0,1.2){17}{
\put(20.00,10.00){\makebox(0,0)[cc]{*}}}
\put(40.00,7.00){\makebox(0,0)[ct]{a}}
\put(91.00,7.00){\makebox(0,0)[ct]{b}}
\bezier{152}(80.00,10.00)(81.00,21.00)(100.00,40.00)
\bezier{140}(85.00,10.00)(87.00,24.00)(100.00,40.00)
\put(100.00,20.00){\line(-1,1){9.67}}
\thicklines
\bezier{84}(100.00,20.00)(90.00,14.00)(81.00,14.00)
\thinlines
\put(84.00,17.00){\circle*{0.00}}
\put(86.00,17.00){\circle*{0.00}}
\put(84.00,19.00){\circle*{0.00}}
\put(86.00,19.00){\circle*{0.00}}
\put(85.00,21.00){\circle*{0.00}}
\put(87.00,21.00){\circle*{0.00}}
\put(86.00,23.00){\circle*{0.00}}
\put(88.00,23.00){\circle*{0.00}}
\put(89.00,27.00){\circle*{0.00}}
\put(91.00,28.00){\circle*{0.00}}
\put(92.00,30.00){\circle*{0.00}}
\put(93.00,32.00){\circle*{0.00}}
\put(94.00,33.00){\circle*{0.00}}
\put(95.00,34.00){\circle*{0.00}}
\put(90.00,26.00){\circle*{0.00}}
\end{picture}

{\noindent \small
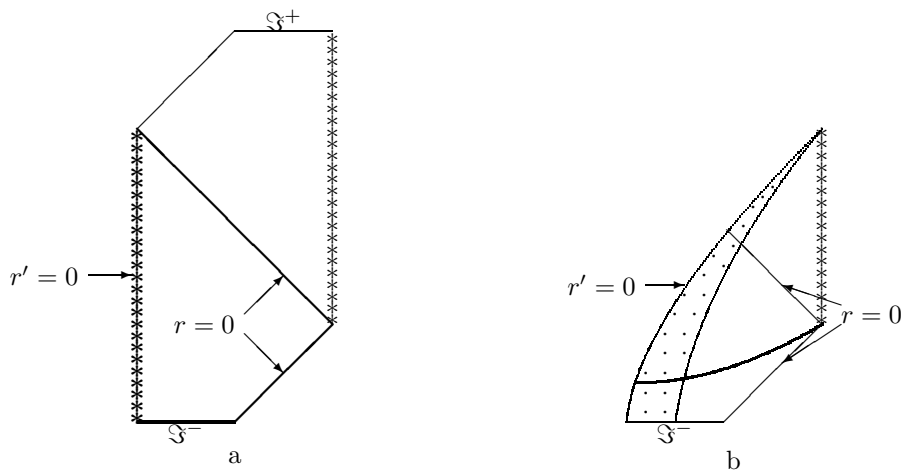
{\bf Fig.\ 10a.} Conformal diagram of history of axis from one black hole
to ``infinity'' for a two-black-hole contracting KT geometry. The black
holes correspond to the region on the right, and the surrounding ``de
Sitter background'' is the region on the left. The total mass is assumed
to be overmassive, so this ``background'' is really an overmassive RNdS
geometry with the singularity shown on the left. Therefore the axis
extends to $r=\infty$ only for $\tau <0$, after that it hits this singularity
at $U(r,\tau)=0$. The $C^1$ extension across the upper $r=0$ horizon
was chosen to be the time-reverse of the heavily outlined region.

{\bf Fig.\ 10b.} In this diagram the cosmological singularity of Fig.\ 10a
is covered up by a sphere of dust, as discussed in Sect.\ 4.1. The dust
region is shown dotted. The thick curve is a spacelike surface with
nonsingular initial data, containing two infinite charged black hole
throats and, on the antipodal point of the universe, a collapsing sphere
of dust. All observers have to cross the future $r=0$ Cauchy horizon and
can thereafter see the naked singularity shown on the right, the result
of the merger of the two black holes. The spacetime ends after a finite
proper time in a ``big crunch.''
}


The region near $r=\infty$ of a KT solution (7) always behaves like an
RNdS geometry with mass equal to the total mass $\sum M_i$. If this
is overmassive, there will be a curvature
singularity in this region, whether the
individual (undermassive) holes have merged or not. This singularity at
the antipodal point of the universe has nothing directly to do with the
black hole merger. To obtain nonsingular initial data we can eliminate
this singularity by replacing it with a collapsing sphere of dust, as in
Sect.\ 4.1. The resulting diagram is shown in Fig.\ 10b. The
initial data induced on the spacelike surface shown by the thick curve
are now nonsingular. In the time development shown, beyond the future
Cauchy horizons $r=0$, a curvature singularity appears. It comes in
from infinity through the infinite throats of the merged black holes
and spreads to $r'=0$, i.e. to the antipodal point of the universe.
The fact that the infinite throats are hidden behind trapped surfaces
does not seem to be sufficient to prevent the singularity from coming
``out of'' the throat. Perhaps a less coordinate-oriented way of saying
this is that all of space collapses down the throats, carrying all
observers with it.

It is clear from the diagram that all observers originating on the initial
surface will reach the Cauchy horizon, and if they extend beyond, they
will see the singularity. We have seen that the Cauchy horizon surrounding
a typical KT throat is not smooth, so that delicate observers may not
survive the crossing. But by distributing several KT masses symmetrically
about a given one, we can make that one as differentiable as necessary
to ensure an observer's survival. So it is reasonable to conclude that
cosmic censorship is violated in these examples.

The initial data in these examples already contain the black holes' infinite
throats, and are not compact. Can we first form the black holes from
collapsing dust, and then let them go through the above scenario? We have
seen in Fig.\ 9 that we can have simultaneous collapse of a dust ball to
form a black hole, and simultaneously remove the overmassive singularity
at the antipodal point by another dust ball. The problem is now that in
the KT solution, as in the RNdS case shown in Fig.\ 9, the two balls
collide before any singularities have formed, at least if the balls move
on $r=0$ resp.\ $r'=0$ trajectories in KT (cosmological) coordinates. Even
if we allow more general trajectories we know for charged test particles that
the trajectories tend to avoid singularities of the same charge; and even
if naked singularities were formed later in the evolution, one would not
know whether they were a fundamental property of the theory, or due to
the dust approximation (``shell crossing singularities,'' which occur
also in the absence of gravity and hence have nothing to do with cosmic
censorship).

Even if we accept the KT solution's infinite throats in place of compact
initial data, we still do not yet have a serious
violation of cosmic censorship, because the general KT solution is still
quite special. The initial position and masses can be specified arbitrarily,
but not the initial velocities. The constraints on the initial values
can be solved in more general (but still not quite generic) contexts,
for example one can drop the $Q_i^2 = M_i^2$ condition \cite{BHKT}.
These initial data can be analytic, but we do not know what happens
beyond the Cauchy horizon. In the general KT solution we have seen that
one has to cross the Cauchy horizon to see the naked singularity. It is
not clear whether more generic solutions have a Cauchy horizon with a
stronger singularity than the KT solution. If so, then cosmic censorship
would be preserved.

\def\refname

\end{document}